\documentclass[12pt]{article}                   

\usepackage{amssymb}
\usepackage{amsmath}
\usepackage{verbatim}
\usepackage{graphicx} 
\usepackage{cite}
\usepackage{geometry}             
\usepackage[hyperindex=true,
          pdfstartview=FitH,
          bookmarksnumbered=true,
          bookmarksopen=true,
          citecolor=blue,
          linkcolor=blue,
          colorlinks=true,
          pdfborder=001,
          unicode]{hyperref}

\newcommand{\rd}{\mathrm{d}}
\newcommand{\pf}[2]{\frac{\partial #1}{\partial #2}}
\newcommand{\pfrac}[2]{\left(\frac{\partial #1}{\partial #2}\right)}
\newcommand{\bfrac}[2]{\left(\frac{#1}{#2}\right)}

\allowdisplaybreaks

\newcommand{\blue}[1]{\textcolor{blue}{#1}}
\renewcommand{\blue}[1]{\textcolor{black}{#1}}  

\begin{document}

\title{Thermodynamics for higher dimensional rotating black holes 
with variable Newton constant}
\author{Liu Zhao 
\\
School of Physics, Nankai University, Tianjin 300071, China
\\{\em email}: \href{mailto:lzhao@nankai.edu.cn}{lzhao@nankai.edu.cn}
}

\date{}

\maketitle

\vspace{5em}

\begin{abstract}
The extensivity for the thermodynamics of general $D$-dimensional rotating black holes 
with or without a cosmological constant can be proved analytically, provided 
the effective number of microscopic degrees of freedom
and the chemical potential are given respectively as $N=L^{D-2}/G, \mu= GTI_D/L^{D-2}$, 
where $G$ is the variable Newton constant, $I_D$ is the Euclidean action and $L$ is a 
constant length scale. In the cases without a cosmological constant, i.e. the Myers-Perry 
black holes, the physical mass and the intensive variables can be expressed as explicit 
macro state functions in the extensive variables in a simple and compact form, 
which allows for an analytical calculation for the heat capacity. The results
indicate that the Myers-Perry black holes with zero, one and $k$ equal rotation
parameters are all thermodynamically unstable.

\end{abstract}

\newpage
\section{Introduction\label{Intro}}

This work is a continuation of the recent works \cite{Zhao1,Zhao2,Zhao3} on black 
hole thermodynamics to the cases of general rotating black holes in $D$-dimensional
Einstein gravity. The formalism introduced in \cite{Zhao1,Zhao2,Zhao3} involves
a variable Newton constant $G$, which enters the expression 
\begin{align}
N = \frac{L^{D-2}}{G}	\label{N}
\end{align}
for the new thermodynamic variable $N$. For asymptotically AdS black holes, $N$ is 
proportional to the central charge of the dual CFT, while for non-AdS black holes, 
$N$ may be simply understood as the effective number of microscopic degrees of freedom
for the black holes. The thermodynamic conjugate of $N$ is given by the chemical potential 
\begin{align}
\mu= \frac{GTI_D}{L^{D-2}}, \label{mu}
\end{align}
where $I_D$ is the Euclidean action and $L$ is a constant length scale (which 
may be identified as the largest radius of the event horizon during a thermodynamic 
process of interest). This formalism coincides neither with the traditional formalism 
\cite{Bekenstein,Bekenstein2,Bardeen,Hawking} nor with the so-called 
extended phase space formalism \cite{Kastor,Dolan,Dolan2,Dolan3,Kubiznak,Cai,Kubiznak2}, 
but is closely related with Visser's holographic thermodynamics \cite{Visser}. 
The idea to introduce a chemical potential in black hole thermodynamics
has already appeared in \cite{Kastor2,Zhang,Karch,Maity,Wei}, 
while a variable Newton constant in black hole thermodynamics has also appeared 
in \cite{Volovik}. However, the major points of concern are completely 
different. The motivation of the works \cite{Zhao1,Zhao2,Zhao3} is mainly to introduce a 
formalism in which the first law and the Euler relation hold simultaneously, which makes the
thermodynamics extensive, and the thermodynamic potential and the intensive variables 
behave as appropriate homogeneous functions in the extensive variables. In this spirit,
the ideas of the works \cite{Tian1,Tian2} are quite close to ours, but with a  
different set of extensive variables. The chemical potential introduced in \cite{Tian1,Tian2} is 
conjugate to the so-called topological charge, while the chemical potential introduced in 
\cite{Zhao1,Zhao2} is conjugate to the central charge of the dual CFT for AdS black holes, 
and the formalism is extended to non-AdS cases without a holographic dual in \cite{Zhao3}. 
The wider applicability of the variable Newton constant formalism may be a signature for its
universality, and one of the purposes of the present work is to illustrate the power and strength
of this formalism in the cases of general rotating black hole solutions in higher 
dimensional Einstein gravity, regardless of the value and sign of the cosmological constant.

The black hole solutions to be analyzed in this paper was first 
obtained in \cite{gilupapo}. The solutions have $k\equiv [(D-1)/2]$
independent rotation parameters $a_i$ in $k$ orthogonal 2-planes. For vanishing cosmological 
constant, the solutions degenerate into Myers-Perry black holes \cite{Myers}. 
By use of the explicit expression for the chemical potential, it will be shown that 
the Hawking-Page (HP) transition \cite{Hawking2} appears only in the asymptotically 
AdS cases and only if the radius of the event horizon approaches the AdS radius. Moreover, 
by introducing the $N,\mu$ variables, the physical mass as well as all
the intensive variables can in principle be expressed as macro state functions in 
the extensive variables with appropriate homogeneity behaviors. These macro state 
functions are made explicit in the cases with vanishing cosmological constant, 
i.e. the Myers-Perry cases. I will also calculate the heat capacities 
analytically and discuss the thermodynamic instabilities for the Myers-Perry black holes.

\section{General rotating black holes in higher dimensions}

The general $D$-dimensional rotating black hole solutions with cosmological constant 
were first obtained in \cite{gilupapo}.  
In Boyer-Linquist coordinates, the metrics are given by 
\begin{align}
\rd s^2 &= - W\, \left(1 - \lambda\, r^2 \right)\, \rd\tau^2
 + \frac{2G m}{U}\Bigl(W\,\rd\tau
 - \sum_{i=1}^k \frac{a_i\, \mu_i^2\, \rd\varphi_i}
  {\Xi_i }\Bigr)^2
 + \sum_{i=1}^k \frac{r^2 + a_i^2}{\Xi_i}\,\mu_i^2\,
    \rd\varphi_i^2 \nonumber\\
&
 + \frac{U\, \rd r^2}{V-2G m}
 + \sum_{i=1}^{k+\epsilon} \frac{r^2 + a_i^2}{\Xi_i}\, \rd\mu_i^2
 + \frac{\lambda}{W\, \left(1 - \lambda\, r^2 \right)}
    \Bigl( \sum_{i=1}^{k+\epsilon} \frac{r^2 + a_i^2}{\Xi_i}
    \, \mu_i\, \rd\mu_i\Bigr)^2 \,,\label{bl}
\end{align}
where $k\equiv [(D-1)/2]$, $\epsilon\equiv (D-1)$ mod 2, 
\begin{align*}
\sum_{i=1}^{k+\epsilon} \mu_i^2 =1\,,
\end{align*}
and 
\begin{align}
W &\equiv \sum_{i=1}^{k+\epsilon} \frac{\mu_i^2}{\Xi_i}\,,\qquad
U \equiv  r^{\epsilon}\, \sum_{i=1}^{k+\epsilon} \frac{\mu_i^2}{r^2 + a_i^2}\,
\prod_{j=1}^k (r^2 + a_j^2)\,,\label{uwline}\\
V &\equiv r^{\epsilon-2}\, (1 -\lambda\,r^2)\,
   \prod_{i=1}^k (r^2 + a_i^2)\,,\qquad \Xi_i\equiv 1 + \lambda\,a_i^2 .
\label{uvw}
\end{align}
The integer $\epsilon$ is known as the evenness number which is 1 for even $D$ and 0 
for odd $D$. The metrics satisfy $R_{\mu\nu}=(D-1)\,\lambda \, g_{\mu\nu}$. 
The choices $\lambda>0,\,\lambda =0,\, \lambda<0$ correspond
to asymptotically de Sitter, flat (i.e. Myers-Perry) and anti-de Sitter cases, respectively.
Moreover, for $\lambda\neq0$, one has $|\lambda|=\ell^{-2}$, where $\ell$ is the (A)dS radius.
The original solutions were presented in unit $G=1$, however, since I will be discussing a
formalism with variable $G$, the explicit $G$-dependence is brought back carefully.

The outer horizon is located at $r=r_+$, where $r_+$ is the largest
root of $V(r)-2G m=0$. Therefore, one has
\begin{align}
m=\frac{V(r_+)}{2G}
=\frac{1}{2G} r_+^{\epsilon-2}\, (1 -\lambda\,r_+^2)\,\prod_{i=1}^k (r_+^2 + a_i^2).
\label{m}
\end{align}  
The surface gravity $\kappa$ and the area $A$ of
the event horizon are given by \cite{gilupapo} 
\begin{align*}
\kappa &= r_+\, (1 -\lambda\,r_+^2)\, \sum_i\frac1{r_+^2 + a_i^2} 
-\frac{2-\epsilon+\epsilon\lambda r_+^2}{2r_+}\,,\\
A &= {\cal A}_{D-2}{r_+^{\epsilon-1}}\, \prod_i \frac{r_+^2 + a_i^2}{\Xi_i}\,,
\end{align*}
where ${\cal A}_{D-2}$ is the volume of the unit $(D-2)$-sphere:
\begin{align*}
{\cal A}_{D-2}  = \frac{2 \pi^{(D-1)/2}}{\Gamma[(D-1)/2]}\,.
\end{align*}
The Hawking temperature and the entropy are then given by
\begin{align}
T&=\frac{\kappa}{2\pi}
=\frac{1}{2\pi}\left[r_+\, (1 -\lambda\,r_+^2)\, \sum_i\frac1{r_+^2 + a_i^2} 
-\frac{2-\epsilon+\epsilon\lambda r_+^2}{2r_+}\right]\,,
\label{Temp}\\
S&=\frac{A}{4G}=\frac{{\cal A}_{D-2}}{4G}
{r_+^{\epsilon-1}}\, \prod_i \frac{r_+^2 + a_i^2}{\Xi_i}\,.
\label{entropy}
\end{align}
The angular velocities, measured relative to a frame that is non-rotating
at infinity, are given by
\begin{align}
\Omega_i = \frac{(1 -\lambda\,r_+^2)\, a_i}{r_+^2 + a_i^2}\,,
\label{Omegaidef}
\end{align}
and the angular momenta are given by
\begin{align}
J_i = \frac{m\, a_i\, {\cal A}_{D-2}}{4\pi\, \Xi_i\,(\prod_j \Xi_j)}.
\label{djdefs}
\end{align}
The physical mass $E$ of the black holes are related to the mass parameter $m$ via
\begin{align}
E = \frac{m\, {\cal A}_{D-2}}{4\pi\, (\prod_j \Xi_j)}\, 
\Big( \sum_{i=1}^k \frac1{\Xi_i} - \frac{1-\epsilon}{2}\Big),
\label{dmassdefsodd}
\end{align}
Finally, the first law of thermodynamics at fixed $G$ reads
\begin{align}
\tilde\rd E = T\, \tilde\rd S + 
\sum_i \Omega_i\, \tilde\rd J_i\,,\label{firstlaw}
\end{align}
where $\tilde\rd$ denotes the total differential taken when $G$ is considered to be a constant.

\blue{Before closing this section, let me remind that, in \cite{Gibbons2}, the black 
hole parameters such as the surface gravity $\kappa$, the area $A$ of the event horizon, 
the physical mass $E$ and the Euclidean action $I_D$ to be used in the next section 
were presented separately for odd and even $D$. Here I found it more convenient 
to rewrite these quantities for generic $D$ in unified form by use of the evenness 
number $\epsilon$. Inserting the corresponding values for $\epsilon$ will recover the original
values of these quantities given in \cite{Gibbons2}.}

\section{Thermodynamics with variable Newton constant}

The Euclidean actions for the black holes described in the preceding section 
are calculated explicitly in \cite{Gibbons2}. After properly restoring the Newton constant, 
the results read
\begin{align}
I_D = \frac{{\cal A}_{D-2}}{
8\pi\, T(\prod_j \Xi_j)}\,\Big( m+\frac{\lambda r_+^{\,\epsilon}}{G} \, 
\prod_{i=1}^k (r_+^2 + a_i^2) \Big).
\label{IDodd}
\end{align}
It was checked \cite{Gibbons2} that $I_D$ obeys the identity
\begin{align}
E -T\, S - \sum_i \Omega_i \, J_i = T\,  I_D .
\label{ID}
\end{align}
Eqs.\eqref{N} and \eqref{mu} implies $\mu N = T I_D$,
thus eq.\eqref{ID} is recognized to be the Euler relation
\begin{align}
E = T\, S + \sum_i \Omega_i \, J_i +\mu N.
\label{Euler}
\end{align}

Now if $G$ is considered as a variable, one has a different total differential, e.g.
\begin{align*}
\rd m &= \tilde\rd m-\frac{m \rd G}{G},\qquad
\rd S = \tilde\rd S-\frac{S \rd G}{G}.
\end{align*}
Moreover, for any function of the form 
\begin{align}
f(m,a_i,r_+)=g(a_i,r_+)\, m,\label{fgm}
\end{align}
one has
\begin{align}
\rd f = m\,\rd g + g\,\rd m
=m\,\tilde \rd g + g\left(\tilde\rd m-\frac{m \rd G}{G}\right)
=\tilde\rd f -f \frac{\rd G}{G}.
\label{dtd}
\end{align}
Meanwhile, from eq.\eqref{N} it follows that
\[
\frac{\rd G}{G}=-\frac{\rd N}{N}.
\]
Thus eq.\eqref{dtd} can also be written as
\begin{align*}
\tilde\rd f = \rd f-f \frac{\rd N}{N}.
\end{align*}
It is important to note that the quantities $E$ and $J_i$ are all proportional to $m$.
For this reason, 
\begin{align}
\rd E &= \tilde\rd E + E \frac{\rd N}{N}\nonumber\\
&=T\left(\rd S- S\frac{\rd N}{N}\right)
+\sum_i \Omega_i\,\left(\rd J_i -J_i\frac{\rd N}{N}\right) +E \frac{\rd N}{N}
\nonumber\\
&=T\rd S + \sum_i \Omega_i\,\rd J_i +\left(E-TS - \sum_i \Omega_i\, J_i\right)\frac{\rd N}{N}
\nonumber\\
&=T\,\rd S + \sum_i \Omega_i\,\rd J_i +\mu\,\rd N, \label{1stL}
\end{align}
where the Euler relation \eqref{Euler} has been used. The analysis does not rely on 
the choice of $\lambda$ and the concrete value of $D$, provided eqs.\eqref{firstlaw} and
\eqref{ID} are valid, and $\lambda$ and $L$ are both kept as constants. 
Eqs.\eqref{Euler} and \eqref{1stL} lay down the fundamental relations 
in our formalism of black hole thermodynamics.

\blue{Please note that the inclusion of the $(\mu, N)$ variables implies that the first law
\eqref{1stL} corresponds to an open thermodynamic system, the corresponding ensemble is 
grand canonical. One can of course consider the case with $N$ fixed, then the first law 
\eqref{1stL} falls back to \eqref{firstlaw}, which corresponds to a closed thermodynamic 
system, or a canonical ensemble. It should be stressed that even in the latter case, the 
variables $(\mu,N)$ are still meaningful and the Euler relation \eqref{Euler} still holds.
Therefore, our formalism is still different from the traditional formalism which is 
governed only by the first law \eqref{firstlaw} and the generalized Smarr relation, 
however in the lack of Euler relation.}

\section{Hawking-Page transitions with negative $\lambda$}

Before dwelling into the detailed analysis for the thermodynamic behaviors, let me first 
make a brief discussion about the possible HP transitions \blue{in either the canonical
or grand canonical ensembles, i.e. regardless of whether $G$ is variable or not}. 

The HP transition \cite{Hawking2} is a particular kind of transition between 
AdS black hole state and a thermal gas state which is characterized by a vanishing 
Gibbs free energy, or equivalently a vanishing chemical potential.

Using the definition \eqref{mu} and eq.\eqref{IDodd}, one has
\begin{align}
\mu=\frac{{\cal A}_{D-2}}{
8\pi\, N(\prod_j \Xi_j)}\,\Big( m+\frac{\lambda r_+^{\,\epsilon}}{G} \, 
\prod_{i} (r_+^2 + a_i^2) \Big).
\label{muodd}
\end{align}
It is evident that $\mu$ can become zero only when $\lambda<0$. The zero appears when 
\begin{align}
m+\frac{\lambda r_+^{\,\epsilon}}{G}\,\prod_{i} (r_+^2 + a_i^2)=0.	
\label{hpeq}
\end{align}
Substituting eq.\eqref{m} into eq.\eqref{hpeq}, one obtains
\begin{align}
\frac{1}{2r_+^2}(1-\lambda r_+^2)+\lambda =0.	
\label{hpeq2}
\end{align}
Writing $\lambda=-\ell^{-2}$, the solution to eq.\eqref{hpeq2} is found to be
\begin{align*}
(r_+)_{\rm HP}=\ell.	
\end{align*}
Therefore the HP transition occurs precisely when the radius of the event 
horizon reaches the AdS radius.

The temperature at which HP transition occurs is known as the HP 
temperature. In the present case, the HP temperature can be expressed analytically 
using the parameters $\ell$ and $a_i$. The result reads
\begin{align*}
T_{\rm HP}= \frac{\ell}{\pi}\,\sum_i\frac1{\ell^2 + a_i^2} 
-\frac{1-\epsilon}{\ell}.
\end{align*}

\section{Physical mass as a macro state function}

The first law \eqref{1stL} and the Euler relation \eqref{Euler} imply that the 
variables $S,J_i,N$ are extensive and their conjugates, $T,\Omega_i,\mu$ are intensive.
This formalism conforms with the standard extensive thermodynamics, therefore one 
naturally expects that the usual practice for analyzing thermodynamic properties of
macroscopic systems should also be applicable here. In particular, the physical mass
and the intensive variables should all be expressible as homogeneous 
macro state functions in the extensive variables $S,J_i$ and $N$. 

\subsection{Generic $\lambda$}

In the cases with generic $\lambda$, one can get from eqs.\eqref{m} and \eqref{entropy}
that
\begin{align}
\prod_{i} \Xi_i=\frac{m {\cal A}_{D-2}r_+}{2S (1 -\lambda\,r_+^2)}.
\label{Xiiodd}
\end{align}
Inserting eq.\eqref{Xiiodd} into \eqref{djdefs}, one has
\begin{align}
J_i=  \frac{a_i\, S (1 -\lambda\,r_+^2)}{2\pi\, r_+(1 + \lambda\,a_i^2)}.
\label{Ji}
\end{align}
Eq.\eqref{Ji} can be viewed as an algebraic equation for $a_i$, whose solution
gives $a_i$ as functions in $S,J_i,r_+$,
\[
a_i=a_i(S,J_i,r_+).
\]
By inserting the functions $a_i(S,J_i,r_+)$ into eq.\eqref{entropy}, 
a very complicated equation for $r_+$ will arise, the solution of which gives a function
\begin{align}
r_+=r_+(S,\mathcal{J}),
\label{rSJ}
\end{align}
where $\mathcal{J}$ denotes the sequence of all $J_i$. 
This in turn implies that $a_i$ are actually functions in $S$ and $\mathcal{J}$, 
because $r_+$ is no longer an independent variable,
\begin{align}
a_i=a_i(S,J_i,r_+(S,\mathcal{J})). \label{aiSJ}
\end{align}
By scaling arguments it can be seen that the functions $r_+(S,\mathcal{J})$ and 
$a_i(S,J_i,r_+(S,\mathcal{J}))$ 
are all zeroth order homogeneous functions in $S,\mathcal{J}$.
Finally, inserting eqs.\eqref{N}, \eqref{rSJ} and \eqref{aiSJ} into 
\eqref{dmassdefsodd},\eqref{Temp}, \eqref{Omegaidef} and \eqref{muodd}, 
the macro state parameters $E,T,\Omega_i,\mu$ can all be expressed as functions in 
$S,\mathcal{J}$ and $N$. 

Although the corresponding functions are very complicated and do not 
worth to be presented here explicitly, some key features can be recognized 
without much difficult. In particular, $E(S,\mathcal{J},N)$ is a homogeneous function of 
the first order, and $T(S,\mathcal{J},N)$, $\Omega_i(S,\mathcal{J},N)$, 
$\mu(S,\mathcal{J},N)$ are homogeneous 
functions of the zeroth order. These homogeneity behaviors are desired for 
the thermodynamic potential and intensive variables in any extensive thermodynamic systems.

\subsection{$\lambda=0$: Myers-Perry cases}

The overwhelming complicated form for the macro state functions $E(S,\mathcal{J},N)$ and 
$T(S,\mathcal{J},N), \,\Omega_i(S,\mathcal{J},N),\, \mu(S,\mathcal{J},N)$ can be avoided 
if one considers only the cases with $\lambda=0$, i.e. the Myers-Perry cases. In such cases, 
one has
\begin{align}
\Xi_i =1,	\label{Xi1}
\end{align}
hence eqs.\eqref{Xiiodd} and \eqref{Ji} become
\begin{align}
&\frac{m\,{\cal A}_{D-2}}{2}=\frac{S}{r_+}, \label{msr}\\
&\frac{a_i}{r_+} =\frac{2\pi J_i}{S}.  \label{JS}
\end{align}
Substituting eqs.\eqref{Xi1} and \eqref{JS} into eq.\eqref{entropy}, one gets
\begin{align}
S&=\frac{N{\cal A}_{D-2}}{4L^{D-2} }\, r_+^{D-2}
\prod_i \left[1+\left(\frac{2\pi J_i}{S}\right)^2\right].
\label{SS}
\end{align}
This is an algebraic equation for $r_+$, whose solution reads
\begin{align}
r_+ = L\left(
\frac{4 S}{{\cal A}_{D-2} N\prod_i \left[1+\left(\frac{2\pi J_i}{S}\right)^2\right]}
\right)^{1/(D-2)}.
\label{r+}
\end{align}
Inserting the above result into eq.\eqref{JS} one gets an expression for $a_i$ as a macro state
function,
\begin{align}
a_i=2\pi L\,\bfrac{J_i}{S}\left(
\frac{4 S}{{\cal A}_{D-2} N\prod_j \left[1+\left(\frac{2\pi J_j}{S}\right)^2\right]}
\right)^{1/(D-2)}.
\label{aims}
\end{align}
Notice that if all $J_i$ are equal to each other, so are $a_i$.
Finally, substituting eqs.\eqref{Xi1} and \eqref{r+} into eq.\eqref{dmassdefsodd},
one obtains an explicit and very compact expression for the physical mass $E$ 
as a macro state function $E(S,\mathcal{J},N)$,
\begin{align}
E(S,\mathcal{J},N)&= (D-2)K N A \prod_{i} B_i, 
\label{Efos}
\end{align}
wherein 
\begin{align*}
K&= \frac{({\cal A}_{D-2})^{1/(D-2)}}{4^{(D-1)/(D-2)}\pi L}
\end{align*}
is a constant factor, and
\begin{align*}
A= \left(\frac{S}{N}\right)^{(D-3)/(D-2)},\quad
B_i= \left[1+\left(\frac{2\pi J_i}{S}\right)^2\right]^{1/(D-2)}.
\end{align*}
It is evident from eq.\eqref{Efos} that the physical mass is proportional to $N$,
with the coefficient of proportionality being a zeroth order homogeneous function
in the extensive variables.

In principle, one can also get the macro state functions 
$T(S,\mathcal{J},N)$, $\Omega(S,\mathcal{J},N)$ and $\mu(S,\mathcal{J},N)$ explicitly 
by substituting eqs.\eqref{Xi1},\eqref{r+}
and \eqref{aims} into appropriate equations presented in Section 2. However, the 
resulting expressions will be somewhat complicated and it needs some efforts to get them 
simplified. For the sake of simplicity, I will proceed in an alternative way, i.e.
by use of the first law \eqref{1stL} and treat the intensive variables as partial 
derivatives of $E$. The results will be presented in the next section.

\section{Equations of states for Myers-Perry black holes}

In this section I will present the explicit form for the macro state functions 
$T(S,\mathcal{J},N)$, $\Omega(S,\mathcal{J},N)$ and $\mu(S,\mathcal{J},N)$ as the 
equation of states (EOS) for Myers-Perry black holes. 

To begin with, it is necessary to write down the partial derivatives of the 
intermediate functions $A(S,N)$ and $B_i(S,J_i)$. These are given as follows,
\begin{align*}
\pf{A}{S}&=\bfrac{D-3}{D-2}\frac{A}{S},\qquad\qquad\qquad
\pf{A}{N}=-\bfrac{D-3}{D-2}\frac{A}{N},\\
\pf{B_i}{S}&=-\frac{2 \pi J_i B_i}{(D-2) \left(S^2+ 2 \pi  J_i S\right)},\quad\,
\pf{B_i}{J_i}=\frac{2 \pi B_i}{(D-2) \left(S+ 2 \pi  J_i \right)}.
\end{align*}
Using these relations one finds
\begin{align*}
\pf{}{S}\left(A \prod_{j} B_j\right) &=
\frac{\chi(S,\mathcal{J})}{(D-2)S} A \prod_j B_j,\\
\pf{}{J_i}\left(A \prod_{j} B_j\right) &=
\frac{2 \pi}{(D-2) \left(S+ 2 \pi  J_i \right)}A \prod_{j} B_j,\\
\pf{}{N}\left(A \prod_{j} B_j\right) &=
\frac{1}{(D-2)N}A \prod_{j} B_j,
\end{align*}
where 
\begin{align}
\chi(S,\mathcal{J})\equiv D-3 - \sum_{i=1}^k \frac{2 \pi J_i }{S+ 2 \pi  J_i}.
\label{chidefs}
\end{align}
Therefore,
\begin{align}
T(S,\mathcal{J},N)&=\pfrac{E}{S}_{\mathcal{J},N}=K\,\bfrac{N}{S} \chi(S,\mathcal{J}) A \prod_j B_j,
\label{TEOS} \\
\Omega_i(S,\mathcal{J},N)&=\pfrac{E}{J_i}_{S,\,\mathcal{J}\backslash J_i,\,N}=
K\,\bfrac{2 \pi N}{S+ 2 \pi J_i }A \,\prod_{j} B_j ,\\
\mu(S,\mathcal{J},N)&=\pfrac{E}{N}_{S,\mathcal{J}}=K A \prod_{j} B_j.
\end{align}

Several remarks are in due here.

1) The explicit EOS allow for a straightforward re-verification
for the Euler relation \eqref{Euler}. Moreover, one can also find some other mass formulae
using the EOS, e.g.
\begin{align}
E&= \frac{D-2}{D-3}\left(TS+\sum_i \Omega_i J_i\right),
\label{smarr} \\
E&= (D-2)\mu N.	
\end{align}
Eq.\eqref{smarr} is already known as the Smarr relation.

2) The chemical potential $\mu(S,\mathcal{J},N)$ is strictly positive, which indicates that 
there is no HP transition in the asymptotically flat cases, 
and that the microscopic degrees of freedom are repulsive. This latter feature may be 
a signature for thermodynamic instability. More confirmative
evidence for the thermodynamic instabilities will be given in the next section by 
analysis of the heat capacity.

3) The condition $T(S,\mathcal{J},N)\geq 0$ requires
\begin{align}
\chi(S,\mathcal{J})= D-3 - \sum_{i=1}^k \frac{2 \pi J_i }{S+ 2 \pi  J_i} \geq 0.
\label{bound}
\end{align}
Since the expression $\frac{2 \pi J_i }{S+ 2 \pi  J_i}$ increases monotonically 
with $J_i$ and approaches the value $1$ as $J_i\to\infty$ with finite $S$, 
and recall that $k=(D-1-\epsilon)/2$, 
one has
\[
\min \chi(S,\mathcal{J})= D-3-\frac{D-1-\epsilon}{2}=\frac{1}{2}(D+\epsilon-5).
\]
For $D<4$, the bound \eqref{bound} can be violated, signifying that the 
angular momentum cannot be too large. 
$D= 4,5$ are critical in the sense that the bound \eqref{bound} can be
at most saturated but not violated. Therefore the existence of extremal black holes 
of the Myers-Perry class can not be excluded by use of the bound \eqref{bound} alone
in these dimensions.
For $D>5$, $T$ is always strictly positive, which excludes the existence of 
extremal Myers-Perry black holes in higher dimensions.

\section{Heat capacity of Myers-Perry black holes}

The explicit form of the EOS allows for an analytical calculation for the heat capacity 
of Myers-Perry black holes. I will be particularly concentrated in the 
heat capacity associated with the macro processes with fixed $\mathcal{J}$ and $N$, i.e.
\begin{align*}
C_{\mathcal{J},N}=T\pfrac{S}{T}_{\mathcal{J},N}.	
\end{align*}
The calculation of the heat capacity $C_{\mathcal{J},N}$ is essentially 
the calculation of the partial derivative
$\pfrac{S}{T}_{\mathcal{J},N}$. This partial derivative cannot be calculated directly because
$S$ has not been written as an explicit function in $T,\mathcal{J}$ and $N$. However, using 
the EOS \eqref{TEOS}, one can calculate its inverse, i.e. $\pfrac{T}{S}_{\mathcal{J},N}$. 
To make the calculation more concise, it is better to start with the partial
derivative of $\chi(S,\mathcal{J})$ defined in \eqref{chidefs} with respect to $S$, which reads
\begin{align*}
\pf{}{S}\chi(S,\mathcal{J})=\sum_i\frac{2\pi J_i}{(S+2\pi J_i)^2}.
\end{align*}
Using the above result, one has
\begin{align*}
\pfrac{T}{S}_{\mathcal{J},N}
&= K \left[\pf{}{S}\bfrac{N}{S} \right]
\chi(S,\mathcal{J}) A \prod_j B_j+K\bfrac{N}{S}\left(\pf{}{S}
\chi(S,\mathcal{J})\right) A \prod_j B_j
\nonumber\\
&\qquad +K\bfrac{N}{S}\chi(S,\mathcal{J}) \pf{}{S}\left(A \prod_j B_j\right)\nonumber\\
&= -\frac{T}{S} +K\bfrac{N}{S}\sum_i\frac{2\pi J_i}{(S+2\pi J_i)^2} A \prod_j B_j
+\frac{\chi(S,\mathcal{J})}{(D-2)S} T\nonumber\\
&=\chi^{-1}(S,\mathcal{J}) \left[\sum_i\frac{2\pi J_i}{(S+2\pi J_i)^2}
+\frac{\chi^2(S,\mathcal{J})}{(D-2)S} -\frac{\chi(S,\mathcal{J})}{S}\right]T.
\end{align*}
Consequently, the heat capacity can be written as
\begin{align}
C_{\mathcal{J},N}&=\frac{T}{\pfrac{T}{S}_{\mathcal{J},N}}
=\chi(S,\mathcal{J}) \left[\sum_i\frac{2\pi J_i}{(S+2\pi J_i)^2}
+\frac{\chi^2(S,\mathcal{J})}{(D-2)S}
-\frac{\chi(S,\mathcal{J})}{S}\right]^{-1}.
\label{hc}
\end{align}

The analytical result \eqref{hc} for the heat capacity makes it possible to analyze
the thermodynamic (in)stability for Myers-Perry black holes in generic dimensions. 
For arbitrary choices of $J_i$, the detailed analysis can still be quite complicated, 
therefore I will proceed only with some simplified cases.

{\bf 1) $J_i=0$ for all $i$, i.e. the Schwarzschild-Tangherlini cases}

In such cases one has
\begin{align*}
\chi(S,\mathcal{J})=D-3,
\end{align*}
and consequently 
\begin{align*}
C_{\mathcal{J},N}&=-(D-2)S <0,
\end{align*}
which shows that the higher dimensional Schwarzschild-Tangherlini black holes are 
thermodynamically unstable. 

{\bf 2) $J_1=J, \,J_i=0$ for all $i\geq 2$, i.e. the cases with a single rotation parameter}

In these cases one has
\begin{align*}
\chi(S,\mathcal{J})&= D-3 - \frac{2 \pi J}{S+ 2 \pi J}>0 \quad {\rm for}\quad 
D\geq 5 \,\,{\rm and}\,\, J<\infty,\\
{C_{\mathcal{J},N}}&={\chi(S,\mathcal{J})} {\mathcal{D}}^{-1}(S,\mathcal{J}),
\end{align*}
where
\begin{align*}
{\mathcal{D}}(S,\mathcal{J})&\equiv \frac{2\pi J}{(S+2\pi J)^2}
+\frac{\chi^2(S,\mathcal{J})}{(D-2)S}-\frac{\chi(S,\mathcal{J})}{S}\nonumber\\
&=-\frac{(D-4)(4\pi J +{S})^2 +(D-2)S^2}{2(D-2) S (2 \pi J+S)^2}<0
\quad {\rm for}\quad D\geq 5.
\end{align*}
Therefore, the higher dimensional Kerr black holes with a single rotation 
parameters always have a negative heat capacity, indicating that
such black holes are thermodynamically unstable.

{\bf 3) $J_i=J\neq 0$ for all $i$, i.e. the cases with $k$ equal rotation parameters}

In these cases, one has
\begin{align}
\chi(S,\mathcal{J})&=D-3-\frac{D-1-\epsilon}{2}\bfrac{2\pi J}{S+2\pi J}
>0 \quad{\rm for}\quad D\geq 5 \,\,{\rm and}\,\, J<\infty,\\
{C_{\mathcal{J},N}}&={\chi(S,\mathcal{J})}  \tilde{\mathcal{D}}^{-1}(S,\mathcal{J}),
\label{hc2}
\end{align}
where
\begin{align*}
\tilde{\mathcal{D}}(S,\mathcal{J})&\equiv \frac{D-1-\epsilon}{2}\frac{2\pi J}{(S+2\pi J)^2}
+\frac{\chi^2(S,\mathcal{J})}{(D-2)S}-\frac{\chi(S,\mathcal{J})}{S}\nonumber\\
&=-\frac{(D-4)\left({D}\pi J + S\right)^2 +(D-2)^2 S^2}
{D(D-2) S (2 \pi  J+S)^2} <0 \quad {\rm for}\quad D\geq 5.
\end{align*}
One thus concludes that for all $D\geq 5$ the heat capacity \eqref{hc2} is always negative, 
indicating that the higher dimensional Myers-Perry
black holes with equal rotation parameters are all thermodynamically unstable.

Before closing this section, let me mention that the negativeness of the heat capacity
of Myers-Perry black holes has already been studied in previous works
using different methods in various limiting cases, see e.g. \cite{Accetta,Dolan4}. 
However, the representation of the heat capacity purely in terms of the 
extensive variables has not been known to my personal knowledge.

\section{Summary and conclusions}

The major achievements and conclusions of the present paper are summarized as follows.

1) The variable Newton constant formalism for black hole thermodynamics holds for
general rotating black hole solutions in higher dimensional Einstein gravity 
with/without a cosmological constant. In this formalism, the first law and the 
Euler relation hold simultaneously, and the physical mass is fully extensive.

2) It can be inferred from the zero of the chemical potential $\mu= \frac{GTI_D}{L^{D-2}}$ 
that the HP transitions appear only in asymptotically AdS cases and only when the 
radius of the event horizon approaches the AdS radius. The HP temperature can be 
expressed analytically in terms of the AdS radius $\ell$ and the rotation parameters $a_i$.

3) For Myers-Perry black holes, the physical mass and the intensive variables can be written 
as explicit functions in the extensive variables, and the results have a remarkable
simple and compact form. The homogeneity behaviors of these macro state functions 
are transparent.

4) The calculation for the heat capacity of Myers-Perry black holes can be carried out 
analytically and can be shown to be always negative in the example cases with zero, one and 
$k$ equal rotation parameters. The results indicate that the corresponding 
black holes are thermodynamically unstable. The thermodynamic instability might also be 
inferred from the strict positivity of the chemical potential in the 
asymptotically flat cases.

The above results provide more evidence for the applicability and the 
strength of the new formalism for black hole thermodynamics proposed in 
\cite{Zhao1,Zhao2,Zhao3}. It is expected that this formalism should also be 
applicable to black holes in extended theories of gravity. Researches in this direction 
are currently undertaken and progresses will be reported soon after.

\section*{Acknowledgement}
This work is supported by the National Natural Science Foundation of 
China under the grant No. 11575088.

\providecommand{\href}[2]{#2}\begingroup
\footnotesize\itemsep=0pt
\providecommand{\eprint}[2][]{\href{http://arxiv.org/abs/#2}{arXiv:#2}}


\begin{thebibliography}{99}

\bibitem{Zhao1} Z. Gao, L. Zhao, ``Restricted phase space thermodynamics for AdS black holes via holography,''
[\eprint{2112.02386}].

\bibitem{Zhao2} Z. Gao, X. Kong, L. Zhao, ``Thermodynamics of Kerr-AdS black holes 
in the restricted phase space,''
[\eprint{2112.08672}].

\bibitem{Zhao3} T. Wang, L. Zhao, ``Black hole thermodynamics is extensive 
with variable Newton constant,''
[\eprint{2112.11236}].

\bibitem{Bekenstein} J. D. Bekenstein, ``Black holes and the second law,'' 
\href{https://link.springer.com/article/10.1007\%2FBF02757029}{\emph{Lett. Nuovo Cim.} 
4 (1972) 737–740}. 

\bibitem{Bekenstein2} J. D. Bekenstein, ``Black holes and entropy,'' 
\href{http://dx.doi.org/10.1103/PhysRevD.7.2333}{\emph{Phys. Rev. D} 7 
(1973) 2333–2346}. 

\bibitem{Bardeen} J. M. Bardeen, B. Carter, and S. W. Hawking, 
``The Four laws of black hole mechanics,''
\href{https://link.springer.com/article/10.1007\%2FBF01645742}{\emph{Commun. Math. Phys.} 
31 (1973) 161–170}.

\bibitem{Hawking} S. W. Hawking, ``Particle Creation by Black Holes,'' 
\href{https://link.springer.com/article/10.1007\%2FBF02345020}{\emph{Commun. Math. Phys.} 
43 (1975) 199–220}. [Erratum: Commun. Math. Phys. 46, 206 (1976)].

\bibitem{Kastor} D. Kastor, S. Ray, J. Traschen, ``Enthalpy and the mechanics of AdS 
black holes,'' \href{https://iopscience.iop.org/article/10.1088/0264-9381/26/19/195011}
{\emph{Class. Quant. Grav.} 26, 195011 (2009)},
[\eprint{0904.2765}].

\bibitem{Dolan} B. P. Dolan, ``The cosmological constant and the black hole 
equation of state,'' \href{https://iopscience.iop.org/article/10.1088/0264-9381/28/12/125020}
{\emph{Class. Quant. Grav.} 28, 125020 (2011)}, 
[\eprint{1008.5023}].

\bibitem{Dolan2} B. P. Dolan, ``Pressure and volume in the first law of black
hole thermodynamics,'' 
\href{https://iopscience.iop.org/article/10.1088/0264-9381/28/23/235017}
{\emph{Class. Quant. Grav.} 28, 235017 (2011)}, 
[\eprint{1106.6260}].

\bibitem{Dolan3} Dolan, ``Compressibility of rotating black holes,'' 
\href{https://doi.org/10.1103/PhysRevD.84.127503}
{\emph{Phys. Rev. D 84}: 127503 (2011)}, [\eprint{1109.0198}]

\bibitem{Kubiznak} D. Kubiznak, R.B. Mann, ``P-V criticality of charged AdS black holes,''
\href{https://link.springer.com/article/10.1007\%2FJHEP07\%282012\%29033}
{\emph{JHEP} 1207, 033 (2012)}, [\eprint{1205.0559}].

\bibitem{Cai} R.-G. Cai, L.-M. Cao, L. Li, and R.-Q. Yang, ``P-V criticality in 
the extended phase space of Gauss-Bonnet black holes in AdS space,''
\href{https://link.springer.com/article/10.1007\%2FJHEP09\%282013\%29005}
{\emph{JHEP} (2013) 005},[\eprint{1306.6233}]. 

\bibitem{Kubiznak2} D. Kubiznak, R. B. Mann, M. Teo, 
``Black hole chemistry: thermodynamics with Lambda,''
\href{https://iopscience.iop.org/article/10.1088/1361-6382/aa5c69}
{\emph{Class. Quantum Grav.} 34 063001 (2017)}, [\eprint{1608.06147}].


\bibitem{Visser} M. R. Visser, ``Holographic thermodynamics requires a chemical 
potential for color,'' [\eprint{2101.04145}].

\bibitem{Kastor2} D. Kastor, S. Ray, J. Traschen, ``Chemical potential in the 
first law for holographic entanglement entropy,'' 
\href{https://link.springer.com/article/10.1007\%2FJHEP11\%282014\%29120}
{\emph{JHEP} (2014) 120}, [\eprint{1409.3521}].

\bibitem{Zhang} J. -L. Zhang, R. G. Cai, H. Yu, ``Phase transition and thermodynamical 
geometry of Reissner-Nordstr\"om-AdS black holes in extended phase space,''
\href{https://journals.aps.org/prd/abstract/10.1103/PhysRevD.91.044028}
{\emph{Phys. Rev. D} 91 (2015) 044028}, [\eprint{1502.01428}].

\bibitem{Karch} A. Karch, B. Robinson, ``Holographic black hole chemistry,''
\href{https://link.springer.com/article/10.1007\%2FJHEP12\%282015\%29073}
{\emph{JHEP} (2015) 1-15}, 
[\eprint{1510.02472}].

\bibitem{Maity} R. Maity, P. Roy, T. Sarkar, ``Black hole phase transitions and the 
chemical potential,'' 
\href{https://linkinghub.elsevier.com/retrieve/pii/S0370269316307407}
{\emph{Phys. Lett. B} 765 (2017) 386–394}, [\eprint{1512.05541}].

\bibitem{Wei} S. W. Wei, B. Liang, Y. X. Liu, ``Critical phenomena and chemical 
potential of a charged AdS black hole,'' 
\href{https://journals.aps.org/prd/abstract/10.1103/PhysRevD.96.124018}
{\emph{Phys. Rev. D} 96 (2017) 124018}, [\eprint{1705.08596}].

\bibitem{Volovik} G. Volovik, 
``Varying Newton constant and black hole to white hole quantum tunneling,''
\href{https://doi.org/10.3390/universe6090133}{\emph{Universe} 2020, 6(9), 133},
[\eprint{2003.10331}].

\bibitem{Tian1} Y. Tian, X.-N. Wu, H. Zhang, ``Holographic Entropy Production,'',
\href{https://link.springer.com/article/10.1007\%2FJHEP10\%282014\%29170}
{\emph{JHEP} 1410:170 (2014)},
[\eprint{1407.8273}].

\bibitem{Tian2} Y. Tian, ``A topological charge of black holes,''
\href{https://iopscience.iop.org/article/10.1088/1361-6382/ab5343}
{\emph{Class. Quantum Grav.} 36 (2019) 245001},
[\eprint{1804.00249}].

\bibitem{gilupapo} G.W. Gibbons, H. L\"u, D.N. Page and C.N. Pope,
``The general Kerr-de Sitter metrics in all dimensions,'' 
\href{}{\emph{J. Geom. Phys. } {53}, 49 (2005)}, [\eprint{hep-th/0404008}].

\bibitem{Myers} R. C. Myers, M. J. Perry, 
``Black holes in higher dimensional spacetimes,''
\href{https://doi.org/10.1016/0003-4916(86)90186-7}{\emph{Ann. Phys.} 172 (1986) 304}.

\bibitem{Hawking2} S. W. Hawking, D.N. Page, ``Thermodynamics of black holes in 
anti-de Sitter space,''
\href{https://link.springer.com/article/10.1007\%2FBF01208266}
{\emph{Commun. Math. Phys.} 87 (1983) 577}.

\bibitem{Gibbons2}
G. W. Gibbons, M. J. Perry, C. N. Pope, ``The first law of thermodynamics for 
Kerr-Anti-de Sitter black holes,'' 
\href{https://iopscience.iop.org/article/10.1088/0264-9381/22/9/002}
{\emph{Class. Quantum Grav.} 22 (2005) 1503}, [\eprint{hep-th/0408217}].

\bibitem{Accetta} F. S. Accetta, M. Gleiser, 
 ``Thermodynamics of higher dimensional black holes,''  
\href{https://linkinghub.elsevier.com/retrieve/pii/0003491687900030}
{\emph{Ann. Phys.} 176 (1987) 278}.

\bibitem{Dolan4} B. P. Dolan, ``On the thermodynamic stability of rotating black holes 
in higher dimensions — a comparison of thermodynamic ensembles,''
\href{https://iopscience.iop.org/article/10.1088/0264-9381/31/13/135012}
{\emph{Class. Quant. Grav.} 31 (2014) 135012},
[\eprint{1312.6810}].

\end{thebibliography}
\end{document}